\documentclass[pre,twocolumn,showpacs,superscriptaddress]{revtex4-1}

\usepackage{amsmath}
\usepackage[xdvi]{graphicx}%
\usepackage{dcolumn}
\usepackage{amsmath}
\usepackage{color}
\usepackage{bm}
\usepackage{subfigure}

\usepackage{array}

\usepackage[utf8]{inputenc}
\usepackage[T1]{fontenc}
\usepackage{ae,aecompl}

\usepackage{color}
\definecolor{darkblue}{rgb}{0,0,0.6}
\definecolor{darkred}{rgb}{0.6,0,0}
\definecolor{darkgreen}{rgb}{0,0.6,0}

\usepackage[colorlinks=true,urlcolor=darkblue,citecolor=darkblue,linkcolor=darkred,hyperfootnotes=false]{hyperref}

\begin{document}

\title{
Finite-time and finite-size scalings in the evaluation of large-deviation functions: \\ Analytical study using a birth-death process}

\author{Takahiro Nemoto}
\email[]{nemoto@math.univ-paris-diderot.fr}
\email[\\]{esteban\_guevarah@hotmail.com}
\email[\\]{vivien.lecomte@univ-paris-diderot.fr}

\affiliation{Laboratoire de Probabilit\'es et Mod\`eles Al\'eatoires, Sorbonne Paris Cit\'e, UMR 7599 CNRS, Universit\'e Paris Diderot, 75013 Paris, France}

\author{Esteban Guevara Hidalgo}
\affiliation{Laboratoire de Probabilit\'es et Mod\`eles Al\'eatoires, Sorbonne Paris Cit\'e, UMR 7599 CNRS, Universit\'e Paris Diderot, 75013 Paris, France}
\affiliation{Institut Jacques Monod, CNRS UMR 7592, Universit\'e Paris Diderot, Sorbonne Paris Cit\'e, F-750205, Paris, France}

\author{Vivien Lecomte}
\affiliation{Laboratoire de Probabilit\'es et Mod\`eles Al\'eatoires, Sorbonne Paris Cit\'e, UMR 7599 CNRS, Universit\'e Paris Diderot, 75013 Paris, France}

\date{\today}

\begin{abstract}
The Giardin\`a-Kurchan-Peliti algorithm is a numerical procedure that uses population dynamics in order to calculate large deviation functions associated to the distribution of time-averaged observables. 
To study the numerical errors of this algorithm, we explicitly devise a stochastic birth-death process that describes the time-evolution of the population-probability. From this formulation, we derive that systematic errors of the algorithm decrease proportionally to the inverse of the population size. 
Based on this observation, we propose a simple interpolation technique for the better estimation of large deviation functions. The approach we present is detailed explicitly in a  two-state model. 
 
\end{abstract}


\maketitle

\section{Introduction}

\begin{figure}[h]
\centering
\includegraphics[width=1\columnwidth]{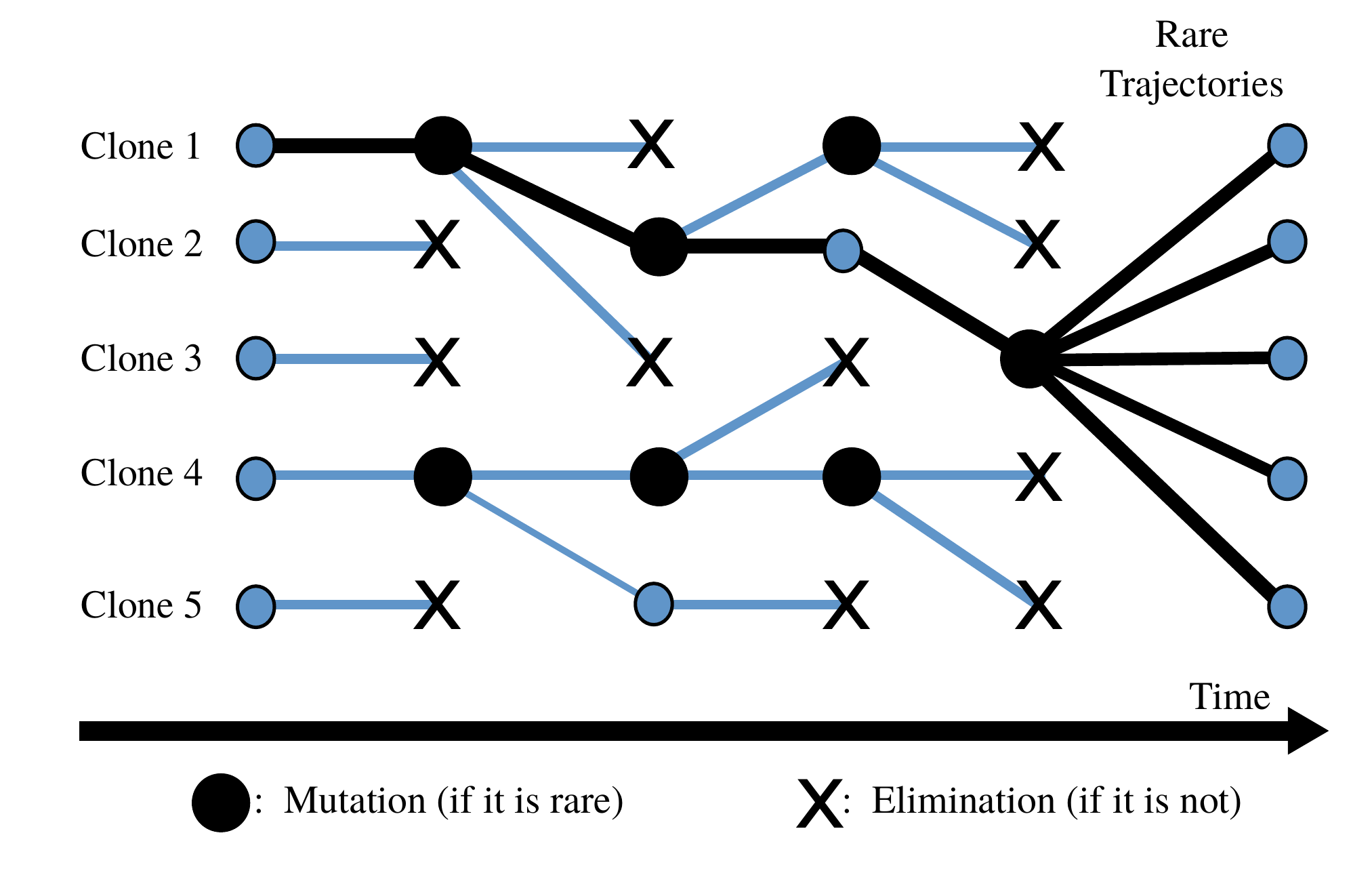}
\caption{ \label{fig:traj}
Schematic picture illustrating the principle of the population dynamics algorithm. `Clones' (or copies) of the system are prepared and they evolve following a mutation-and-selection process, maintaining the total population constant.}
\end{figure}

Cloning algorithms are numerical procedures aimed at simulating rare events efficiently, using a  population dynamics scheme. In such algorithms, copies of the system are evolved in parallel and the ones showing the rare behavior of interest are multiplied iteratively~\cite{Anderson1975_63_4,Glasserman_1996,2001IbaYukito,GRASSBERGER200264,
4117599,CappeGuillinetal,Forwardinterfacesampling,PhysRevLett.94.018104,
delmoral2005,Dean2009562,GuyaderArnaudetal,giardina_direct_2006,lecomte_numerical_2007,
JulienNaturephys_2007,tailleur_simulation_2009,giardina_simulating_2011,
1751-8121-46-25-254002_2013,1751-8121-49-20-205002_2016} (See Fig.~\ref{fig:traj}). 
One of these algorithms proposed by Giardin\`a {\it et al.}~\cite{giardina_direct_2006,lecomte_numerical_2007,
JulienNaturephys_2007,tailleur_simulation_2009,giardina_simulating_2011,
1751-8121-46-25-254002_2013,1751-8121-49-20-205002_2016} is used to evaluate numerically the cumulant generating function (a large deviation function, LDF) of additive (or ``time-extensive'') observables in Markov processes~\cite{opac-b1093895,Touchette20091}. 
It has been applied to many physical systems, including chaotic systems, glassy dynamics and non-equilibrium lattice gas models, and it has allowed the study of novel properties, such as the behavior of breathers in the Fermi-Pasta-Ulam-Tsingou chain~\cite{JulienNaturephys_2007}, dynamical phase transitions in kinetically constrained models~\cite{garrahanjacklecomtepitardvanduijvendijkvanwijland}, and an additivity principle for simple exclusion processes~\cite{PhysRevLett.92.180601,hurtado_large_2010}.

While the method has been used widely, there have been fewer studies focusing on the analytical justification of the algorithm. Even though it is heuristically believed that the LDF estimator converges to the correct result as the number of copies $N_c$ increases, there is no proof of this convergence. Related to this lack of the proof, although we use the algorithm by assuming its validity, we do not have any clue how fast the estimator converges as $N_c \rightarrow \infty$.  
In order to discuss this convergence, we define two types of numerical errors.
First, for a fixed finite $N_c$, averaging over a large number of realizations, the LDF estimator converges to an incorrect value, which is different from the desired large deviation result. We call this deviation from the correct value, {\it systematic errors}. Compared with these errors, we also consider the fluctuations of the estimated value. More precisely, for a fixed value of $N_c$, the results obtained in different realizations are distributed around this incorrect value. We call the errors associated to these fluctuations the {\it stochastic errors}. Although both errors are important in numerical simulations, the former one can lead this algorithm to produce wrong results. For example as seen in Ref.~\cite{NemotoBouchetLecomteJack}, the systematic error grows exponentially as a temperature decreases (or generically in the weak noise limit of diffusive dynamics).

%%%%%%%

In order to study these errors, we employ a birth-death process~\cite{tagkey2007iii,opac-b1079113} description of the population dynamics algorithm as explained below: We focus on physical systems described by a Markov dynamics~\cite{giardina_direct_2006,giardina_simulating_2011,lecomte_numerical_2007} with a finite number of states $M$, and we denote by $i$ ($i=0,1,\cdots M-1$) the states of the system. This Markov process has its own stochastic dynamics, described by the transition rates $w(i\rightarrow j)$.
In population dynamics algorithms, in order to study its rare trajectories, one prepares $N_c$ copies of the system, and simulate
these copies according to {\it (i)} the dynamics of $w(i\rightarrow j)$ (followed independently by all copies) and {\it (ii)}  `cloning' step in which the ensemble of copies is directly manipulated, \emph{i.e.}, some copies are eliminated while some are multiplied (See Table~\ref{tablecorrespondence_}). 
Formally, the population dynamics represents, for a \emph{single} copy of the system, a process that does not preserve probability. This fact has motivated the studies of auxiliary processes~\cite{jack_large_2010}, effective processes~\cite{1742-5468-2010-10-P10007} and  
driven processes~\cite{PhysRevLett.111.120601} to construct modified dynamics (and their approximations~\cite{PhysRevLett.112.090602}) that preserve probability.  
Different from these methods, in this article, we formulate explicitly the meta-dynamics of the copies themselves by using a stochastic birth-death process. The process preserves probability, and it allows us to study the numerical errors of the algorithm when evaluating LDF.

\begin{table*}
\begin{center}
          \caption{\label{tablecorrespondence_} Correspondence between the population dynamics and the birth-death process to describe it.}
          \begin{tabular}{c||c|c}
\vphantom{\Big|}

                     	  			& Population dynamics algorithm      	&        Birth-death process describing    					 \\[-1mm]
		   	  			&								&   the population dynamics  		\\[1mm] \hline \hline
\vphantom{\Big|}
    State of the system	 	 & 		$i$              				  &        $n=(n_0,n_2,\cdots,n_{M-1})$ 					 \\
                       				 &  		($i=0,1,\cdots M-1$)		  &        ($0\leq n_i \leq N_c$ with $\sum_{i}n_i=N_c$)          \\[1mm]  \hline
\vphantom{\Big|}
 Transition rates			&  $w(i \rightarrow j)$ 	   			&	 $\sigma(n \rightarrow \tilde n)$ 					 \\[-1mm]
			 			&  Markov process on states $i$ 	   			&	 Markov process on states $n$						 \\[1mm]  \hline
\vphantom{\Big|}
         Numerical procedure     &	Prepare $N_c$ clones  and evolve those 		&   	Described by the dynamics 					\\[-1mm]
	for rare-event sampling	&    with a mutation-selection procedure  &	 	of rates $\sigma(n \rightarrow \tilde n)$			  \\[1mm] \hline
          \end{tabular}  
        \end{center}
 \end{table*}

In this article,  we consider the dynamics of the copies as a stochastic birth-death process whose state is denoted $n=(n_0,n_1,n_2,\cdots, n_{M-1})$, where $0\leq n_i\leq N_c$ represents the number of copies which are in state $i$ in the ensemble of copies.
We explicitly introduce the transition rates describing the dynamics of $n$, which we denote by $\sigma (n\rightarrow \tilde n)$. We show that the dynamics described by these transition rates lead in general to the correct LDF estimation of the original system $w(i\rightarrow j)$ in the $N_c \rightarrow \infty$ limit. 
We also show that the systematic errors are of the order $\mathcal O(1/N_c)$, whereas the numerical errors are of the order $\mathcal O(1/(\tau N_c))$ (where $\tau$ is an averaging duration). This result is in clear contrast with standard Monte-Carlo methods, where the systematic errors are always 0. Based on this convergence speed, we then propose a simple interpolation technique to make the cloning algorithm more reliable. 
Furthermore, the formulation developed in this paper provides us the possibility to compute exactly the expressions of the convergence coefficients, as we do in Sec.~\ref{Section:Demonstrations} on a simple example.

The analytical analysis presented in this paper is supplemented with a thorough numerical study in a companion paper~\cite{SecondPart}. In the companion paper, we employ an intrinsically different cloning algorithm, which is the continuous-time population dynamics algorithm, that cannot be studied by the methods presented in this paper (see Sec.~\ref{subsubsec:differenceContinuousTime}). We show in the companion paper \cite{SecondPart} that the validity of the scaling that we derive analytically here is very general.
In particular, we demonstrate in practice the efficiency of the interpolation technique in the evaluation of the LDF,
irrespective of the details of the population dynamics algorithm. % even though the numerical study is performed in a different continuous-time settings. 

The construction of this paper is as follows. We first define the LDF problem in the beginning of Sec.~\ref{sec:Birth-deathprocess}, and then formulate the birth-death process used to describe the algorithm in Sec.~\ref{subsec:TransitionmatrixAlgorithm}. By using this birth-death process, we demonstrate that the  estimator of the algorithm converges to the correct large deviation function in Sec.~\ref{Subsec:DerivationLargeDeviationEstimator}. At the end of this section, in Sec.~\ref{subsec:systemsizeexpansion}, we discuss the convergence speed of this estimator (the systematic errors) and derive its scaling $\sim 1/N_c$. In Sec.~\ref{Sec:largedeviation_largedeviation}, we turn to  stochastic errors. For discussing this, we introduce the large deviation function of the estimator, from which we derive that the convergence speed of the stochastic errors is proportional to $1/(\tau N_c)$. In the next section, Sec.~\ref{Section:Demonstrations}, we introduce a simple two-state model, to which we  apply the formulations developed in the previous sections. We derive the exact expressions of the systematic errors in Sec.~\ref{subsection:systematicerrors_twostate} and of the stochastic errors in Sec.~\ref{subsection:largedeviationsInPopTwoState}. At the end of this section, in Sec.~\ref{subsec:Different large deviation estimator}, based on these exact expressions, we propose another large deviation estimator defined in the population dynamics algorithm. In the final section, Sec.~\ref{sec:discussion}, we first summarize the result obtained throughout this paper, and then in Sec.~\ref{subsec:Interporationtechnique}, we propose a simple interpolation technique based on the convergence speed of the systematic errors which allows us to devise a better practical evaluation of the LDF. Finally in Sec.~\ref{subsec:openquestions}, we discuss two open questions.

\section{Birth-Death Process Describing the Population Dynamics Algorithm}
\label{sec:Birth-deathprocess}

%%% State space
As explained in the introduction (also see Table~\ref{tablecorrespondence_}), the state of the population is $n=(n_0,n_1,\cdots,n_{M-1})$, where $n_i$ represents the number of clones in the state $i$. The total population is preserved: $\sum_i n_i = N_c$.  Below, we introduce the transition rates of the dynamics between the occupations $n$, $\sigma(n\rightarrow \tilde n)$ that describe corresponding large deviations of the original system, where the dynamics of the original system is given by the rates $w(i\rightarrow j)$ as detailed below.

As the original system, 
 we consider the continuous-time Markov process in a discrete-time representation. 
By denoting by $dt$ the time step, 
the transition matrix $R_{j,i}$ for time evolution of the state $i$ is described as
\begin{equation}
R_{j,i} = \delta_{i,j} + dt \Big [ w(i\rightarrow j) -\delta_{i,j}\sum_{k}w(i\rightarrow k) \Big ],
\label{eq:ratesRij}
\end{equation}
where we set $w(i\rightarrow i)=0$. The probability distribution of the state $i$, $p_i(t)$, evolves in time as $p_i(t+dt) = \sum_{j} R_{i,j}p_j(t)$. In the $dt\rightarrow 0$ limit, one obtains the continuous-time master equation describing the evolution of $p_i(t)$~\cite{tagkey2007iii,opac-b1079113}. For simplicity, especially for the cloning part of the algorithm, we keep here a small finite $dt$. 
The reason why we use a discrete-time representation is solely for simplicity of the discussion. The main results can be derived even if we start with a continuous-time representation (see Sec.~\ref{subsubsec:dtDeltat}).
%
% For the cloning process however, following the original proposal of Giardin\`a, Kurchan and Peliti~\cite{giardina_direct_2006}, the time discretization facilitates considerably the analytical study of the algorithm, by bounding the number of offsprings that an individual can have at each cloning step (see).
%
For the original dynamics described by the transition matrix~\eqref{eq:ratesRij}, we consider
an observable  $b_i$ depending on the state $i$ and we are interested in the distribution of its time-averaged value during a time interval $\tau$, defined as
\begin{equation}
B(\tau ) = \frac{1}{\tau} \sum_{t=0}^{\tau / dt} dt \ b_{i(t)}.
\label{eq:defBtau}
\end{equation}
Here $i(t)$ is a trajectory of the system generated by the Markov dynamics described by $R_{j,i}$. We note that $B(\tau)$ is a path- (or history-, or realization-) dependent quantity. Since $\tau B(\tau)$ is an additive observable, the fluctuations of $B(\tau)$ depending on the realizations are small when $\tau$ is large, but one can describe the large deviations of $B(\tau)$. Those occur with a small probability, and obey a large deviation principle. We denote by ${\rm Prob}(B)$ the distribution function of $B(\tau)$. The large deviation principle ensures that ${\rm Prob}(B)$ takes an asymptotic form  ${\rm Prob}(B) \sim \exp (- \tau I(B))$ for large $\tau$, where $I(B)$ is a large deviation function (or `rate function')~\cite{Touchette20091,opac-b1093895}. 
If the rate function $I(B)$ is convex, the large deviation function is expressed as a Legendre transform of a cumulant generating function (CGF) $\psi(s)$ defined as
\begin{equation}
\psi(s) = \lim_{\tau \rightarrow \infty} \frac{1}{\tau} \log \left \langle e^{-s \tau B(\tau)} \right \rangle,
\label{eq:defpsi}
\end{equation}
namely: $I(B) = - \inf_{s}\left [ s B +  \psi(s)  \right ]$. The large deviation function $I(B)$ and this generating function $\psi(s)$ are by definition difficult to evaluate numerically in Monte-Carlo simulations of the original system of transition rates $w(i\rightarrow j)$ (see, for example,~\cite{PhysRevE.92.052104}). To overcome this difficulty, population dynamics algorithms have been developed~\cite{giardina_direct_2006,lecomte_numerical_2007,
JulienNaturephys_2007,tailleur_simulation_2009,giardina_simulating_2011,
1751-8121-46-25-254002_2013,1751-8121-49-20-205002_2016}.
Here, we describe this population dynamics algorithm by using a birth-death process on the occupation state $n$ allowing us to study  systematically the errors in the estimation of $\psi(s)$ within the population dynamics algorithm. We mention that, without loss of generality, we restrict our study to so-called `type-B' observables that do not depend on the transitions of the state~\cite{garrahan_first-order_2009}, \emph{i.e.}~which are time integrals of the state of the system, as in~\eqref{eq:defBtau}. Indeed, as explained for example in Refs.~\cite{giardina_simulating_2011} and~\cite{NemotoBouchetLecomteJack}, one can always reformulate the determination of the CGF of mixed-type observables into that of a type-B variable, by modifying the transition rates of the given system.

\subsection{Transition Matrices Representing the Population Dynamics Algorithm}
\label{subsec:TransitionmatrixAlgorithm}

We denote the probability distribution of the occupation~$n$ at time~$t$ by $P_n(t)$. The time-evolution of this probability is decomposed into three parts.
 The first one is the original Monte-Carlo dynamics based on the transition rates $w(i\rightarrow j)$. The second one is the cloning procedure of the population dynamics algorithm, which favors or disfavors configurations according to a well-defined rule. The third one is a supplementary (but important) part which maintains the total number of clones to a constant $N_c$. We denote the transition matrices corresponding to these steps by $\mathcal T$, $\mathcal C$ and $\mathcal K$, respectively. By using these matrices, then, the time evolution of the distribution function is given as
\begin{equation}
P_{n}(t+dt) = \sum_{ \tilde n } \left (  \mathcal  K \mathcal C \mathcal T  \right )_{n,\tilde n} P_{\tilde n}(t).
\label{eq:time_Evolution}
\end{equation}
We derive explicit expressions of these matrices in the following sub-sections. 
We also summarize the obtained results in Table~\ref{table_Matrices}.

\begin{table*}
\begin{center}
          \caption{\label{table_Matrices} Transition matrices (see Eq.~(\ref{eq:time_Evolution})) describing the birth-death process.}

          \begin{tabular}{c||c}
\vphantom{\bigg|}
                     	  			& Transition matrices      					 \\[2mm]  \hline \hline
\vphantom{\bigg|}
    Dynamics (``mutations'')	 	 & 		$ \mathcal T_{\tilde n, n} \equiv   \delta_{\tilde n, n} +  dt    \sum_{i=0}^{M-1} n_i  \sum_{j=0, (j\neq i)}^{M-1}w(i \rightarrow  j)    \left [  \delta _{\tilde n_i,n_i-1}   \delta _{\tilde n_j,n_j+1} \ \delta ^{i,j}_{\tilde n,n} \  -  \delta _{\tilde n,n}    \right ]  $   \\[2mm]  \hline
\vphantom{\bigg|}
  Cloning (``selection'')		&  $\mathcal C_{\tilde n, n}  =   \delta_{\tilde n, n} + s \ dt    \sum_{i=0}^{M-1}   n_i |\alpha_i|   \left [  \delta _{\tilde n_i,n_i + \alpha_i/|\alpha_i|}  \  \delta^{i}_{\tilde n, n}  - \delta_{\tilde n, n}   \right ]      + \mathcal O(dt^2)$            \\[2mm]  \hline
\vphantom{\bigg|}
  Maintaining $N_c$  	&   $\mathcal K_{\tilde n,n}  =  \delta_{\sum n_i,N_c}  \delta_{\tilde n, n}  
 + \sum_{k=-1,1} \delta_{\sum_i n_i, N_c + k}  \sum_{i=0}^{M-1}  \delta_{\tilde n_i, n_i - k} \  \delta^{i}_{\tilde n, n} \  \frac{n_i }{N_c + k}   $  \\[2mm] \hline
\vphantom{\bigg|}
	 Full process 		& $   (\mathcal K \mathcal C \mathcal T)_{\tilde n, n} 
  =   \delta_{\tilde n, n} +  dt    \sum_{i=0}^{M-1} n_i  \sum_{j=0, (j\neq i)}^{M-1} \left [ w(i \rightarrow  j) + s \tilde w_{n}(i \rightarrow  j) \right ]  \left [  \delta _{\tilde n_i,n_i-1}   \delta _{\tilde n_j,n_j+1} \ \delta ^{i,j}_{\tilde n,n} \  -  \delta _{\tilde n,n}    \right ]$  \\[2mm] 
				&  with  \ \ \  $\tilde w_n (i \rightarrow  j) = \frac{n_j  }{N_c} \left [ \alpha_j  \delta_{j\in \Omega^{(+)}} \frac{N_c}{N_c + 1} - \alpha_i \delta_{i\in \Omega^{(-)}}  \frac{N_c}{N_c - 1} \right ]$            			\\[2mm] \hline
          \end{tabular}  
        \end{center}
 \end{table*}

\subsubsection{Derivation of the Original Dynamics Part, $\mathcal T$}

We first consider the transition matrix $\mathcal T$, which describes the evolution of the occupation state $n$ solely due to the dynamics based on the rates $w(i\rightarrow j)$. During an infinitesimally small time step $dt$, the occupation $n=(n_0,n_1,\cdots,n_{M-1})$ changes to $\tilde n = (n_0,n_1,\cdots, n_i-1, \cdots, n_j+1, \cdots, n_{M-1})$ where $0\leq i< M$ and $0\leq j< M$ (for all $i\neq j$). Since there are $n_i$ clones in the state $i$ before the transition, the transition probability of this change is given as $n_i w(i\rightarrow j) dt$. Thus, we obtain
\begin{equation}
\begin{split}
 \mathcal T_{\tilde n, n} \equiv   \delta_{\tilde n, n} + & dt    \sum_{i=0}^{M-1} n_i  \sum_{j=0, (j\neq i)}^{M-1}w(i \rightarrow  j)   \\ 
& \times \left [  \delta _{\tilde n_i,n_i-1}   \delta _{\tilde n_j,n_j+1} \ \delta ^{i,j}_{\tilde n,n} \  -  \delta _{\tilde n,n}    \right ],
\end{split}
\end{equation}
where $ \delta ^{i,j}_{\tilde n,n}$ is a Kronecker-delta for the indices except for $i,j$: $\delta ^{i,j}_{\tilde n,n} \equiv \prod_{k\neq i,j} \delta_{\tilde n_k,n_k}$. 
One can easily check that this matrix satisfies the conservation of the probability: $\sum_{\tilde n} \mathcal T_{\tilde n, n}=1$.
It corresponds to the evolution of $N_c$ independent copies of the original system with rates $w(i\to j)$.

\subsubsection{Derivation of the Cloning Part, $\mathcal C$}
\label{subsubsec:CloningRatio}

In the population dynamics algorithm (for example the one described in the Appendix A of Ref.~\cite{NemotoBouchetLecomteJack}), at every certain time interval $\Delta t$, 
one evaluates the exponential factor for all clones, which is equal to $e^{-s\int_t ^{t+\Delta t}\! dt' ~ b_{i(t')}}$ if the clone is in state $(i(t'))_{t'=t}^{t+\Delta t}$ during a time interval $t\leq t' \leq t +\Delta t$. We also call this exponential factor {\it cloning ratio}, because  % (or the renormalized one $e^{-s ~ \int _t^{t+\Delta t} dt' ~ b_{i(t')}}/\sum_{{\rm clones}}e^{-s ~\int_{t}^{t+\Delta t} dt' ~ b_{j(t')}}$, see~\cite{NemotoBouchetLecomteJack} for example). 
this factor determines whether each clone is copied or eliminated after this time interval.
Although the details of how to determine this selection process can depend on the specific type of algorithms, the common idea is that each of the clones is copied or eliminated in such a way that a clone in state $i(t)$ has a number of descendant(s) proportional to the cloning factor
%$e^{-s \int_t ^{t+\Delta t}\! dt' ~ b_{i(t')}}$ 
on average after this time interval.

In order to implement this idea in our birth-death process, we assume this time step $\Delta t$ to be small. For the sake of simplicity, we set this $\Delta t$ to be our smallest time interval $d t$: $\Delta t =dt$. This condition is not mandatory whenever the $\Delta t \rightarrow 0$ limit is taken at the end (see Sec.~\ref{subsubsec:dtDeltat} for the case $\Delta t > dt$). Then, noticing that the time integral $\int_t ^{t+\Delta t} dt' ~ b_{i(t')}$ is expressed as $ dt \ b_{i(t)}$ for small $dt$, we introduce the following quantity for each state $i$ ($i=0,1,2,...,M-1$):
\begin{equation}
\nu_i \equiv \frac{ n_i e^{-s ~ dt ~ b_i} }{\sum_{j=0}^{M-1} n_j e^{-s ~ dt ~ b_j  }  }  N_c.
\label{eq:nufac}
\end{equation}
Note that there is a factor $n_i$ in front of the exponential function $e^{-s ~ dt ~ b_i}$
 which enumerates the number of clones that occupy the state $i$. The quantity $\nu_i$ is aimed at being the number of clones in state $i$ after the cloning process, however, since $\nu_i$ is not an integer but a real number, one needs a supplementary prescription to fix the corresponding integer number of descendants. In general, in the implementation of population dynamics, this integer is generated randomly from the factor $\nu_i$, equal either to its lower or to its upper integer part. 
The probability to choose either the lower or upper integer part is fixed by imposing that the number of descendants is equal to $\nu_i$ on average.
For instance, if $\nu_i$ is equal to $13.2$, then $13$ is chosen with probability $0.8$, and $14$ with probability $0.2$. 
Generically, $\lfloor \nu_i \rfloor$
and $\lfloor \nu_i \rfloor + 1$ are chosen with probability $1+ \lfloor \nu_i  \rfloor - \nu_i$ and $ \nu_i - \lfloor \nu_i  \rfloor $, respectively.
We note that we need to consider these two possibilities for all indices $i$.
We thus arrive at the following matrix:
\begin{equation}
\begin{split}
\mathcal C_{\tilde n, n} \equiv  &  \sum_{x_0 = 0}^1  \sum_{x_1 = 0}^1  \sum_{x_2 = 0}^1 \cdots  \sum_{x_{M-1} = 0}^1  \prod_{i=0}^{M-1}   \\
& \times \delta _{\tilde n_i, \lfloor \nu_i  \rfloor  + x_i} \left [ \left ( \nu_i - \lfloor \nu_i  \rfloor \right ) x_i +  \left (  1+ \lfloor \nu_i  \rfloor - \nu_i \right ) (1 - x_i) \right ].
\label{eq:transitionMatrixForCopyDef}
\end{split}
\end{equation}

Now, we expand $\mathcal C$ at small $dt$ and we keep only the terms proportional to $\mathcal O(1)$ and $\mathcal O(dt)$, which do not vanish in the continuous-time limit. For this purpose, we expand $\nu_i$ as
\begin{equation}
\nu_i = n_i \bigg [ 1 + s \ dt \Big ( \sum_j \frac{n_jb_j}{N_c} - b_i \Big ) \bigg ] + \mathcal O(dt^2),
\end{equation}
where we have used $\sum_i n_i =N_c$.
This expression indicates that $\lfloor \nu_i \rfloor$ is determined depending on the sign of $\sum_j n_jb_j / N_c - b_i $, where we assumed $s>0$ for simplicity without loss of generality (because when $s<0$, we can always re-define $-b$ as $b$ to make $s$ to be positive).  By denoting this factor by $\alpha_i$, \emph{i.e.}
\begin{equation}
\alpha_i(n) \equiv \sum_j \frac{n_jb_j}{  N_c} - b_i,  
\label{def:eqalphai}
\end{equation}
we thus define the following state-space $\Omega^{(\pm)}(n)$:
\begin{equation}
 \Omega^{(\pm)}(n)   = \left \{ \  i \  \big | \ 0\leq i < M  \ {\rm and} \    \pm \alpha_i(n)  > 0 \right  \}.
\end{equation}
From this definition, for sufficiently small $dt$, we obtain
\begin{equation}
\left \lfloor \nu_i \right \rfloor = n_i
\end{equation}
for $i  \in \Omega^{(+)}$, and
\begin{equation}
\left \lfloor \nu_i \right \rfloor = n_i - 1
\end{equation}
for $i  \in \Omega^{(-)}$. Substituting these results into (\ref{eq:transitionMatrixForCopyDef}) and expanding in $dt$, we obtain (denoting here and thereafter $\alpha_i=\alpha_i(n))$:
\begin{equation}
\begin{split}
\mathcal C_{\tilde n, n}  = &  \delta_{\tilde n, n} + s \ dt    \sum_{i=0}^{M-1}   n_i |\alpha_i|   \big [  \delta _{\tilde n_i,n_i + \alpha_i/|\alpha_i|}  \  \delta^{i}_{\tilde n, n}  - \delta_{\tilde n, n}   \big ]   \\
&   + \mathcal O(dt^2),
\label{eq:Cdt0limit}
\end{split}
\end{equation}
where $\delta^{i}_{\tilde n, n}$ is a Kronecker delta for the indices except for $i$: $\delta^{i}_{\tilde n, n} = \prod_{k\neq i}\delta_{\tilde n_k,n_k}$. One can easily check that this matrix preserves probability: $\sum_{\tilde n} \mathcal C_{\tilde n, n} = 1$.

\subsubsection{Derivation of the Maintaining Part, $\mathcal K$}

As directly checked, the operator $\mathcal T$ preserves the total population $\sum_i n_i$.
However, the operator representing the cloning $\mathcal C$, does not.
In our birth-death implementation, 
this property originates from the rounding process $\lfloor \nu_i  \rfloor$ in the definition of $\mathcal C$:
even though $\nu_i$ itself satisfies $\sum_i \nu_i = N_c$, because of the rounding  process of $\nu_i$, the number of clones after multiplying by $\mathcal C$ (that is designed to be proportional to $\nu_i$ on average) can change.
There are several ways to keep the number  $N_c$ of copies constant without biasing the distribution of visited configurations. One of them is to choose randomly and \emph{uniformly} $\delta N_c$ clones from the ensemble, where $\delta N_c$ is equal to the number of excess (resp.~lacking) clones with respect to $N_c$, and to eliminate (resp.~multiply) them.

In our birth-death description, we implement this procedure as follows. We denote by $\mathcal K$ the transition matrix maintaining the total number of clones to be the constant~$N_c$. 
We now use a continuous-time asymptotics $dt\rightarrow 0$. In this limit, from the expression of the transition matrix elements~(\ref{eq:Cdt0limit}), we find that at each cloning step the number of copies of the cloned configuration varies by $\pm 1$ at most. Hence, the total number of clones after multiplying by $\mathcal C$, $\sum_{i} n_i$, satisfies the following inequality
\begin{equation}
N_c - 1 \leq  \sum_{i}  n_i  \leq  N_c +1.
\end{equation}
Among the configurations $n$ that satisfy this inequality, there are three possibilities, which are $\sum_{i}n_i=N_c$ and $\sum_{i}n_i=N_c\pm 1$. 
If $n$ satisfies $\sum_{i}n_i=N_c$, we do not need to adjust $n$, 
while if $n$  satisfies
$\sum_{i}n_i=N_c + 1$ (resp.~$\sum_{i}n_i=N_c - 1$), we eliminate (resp.~multiply) a clone chosen randomly and uniformly. Note that, in our formulation, we do not distinguish the clones taking the same state. This means that we can choose one of the occupations $n_i$ of a state $i$ according to a probability proportional to the number of copies $n_i$ in this state. In other words, the probability to choose the state $i$ and to copy or to eliminate a clone from this state is proportional to $ n_i  / \sum_{j=0}^{M-1}n_j $. Therefore, we obtain the expression of the matrix $\mathcal K$ as
\begin{equation}
\begin{split}
\mathcal K_{\tilde n,n}  = & \delta_{\sum_i\! n_i,N_c}  \delta_{\tilde n, n}  \\
& + \sum_{k=-1,1} \delta_{\sum_i\!n_i, N_c + k}  \sum_{i=0}^{M-1}  \delta_{\tilde n_i, n_i - k} \  \delta^{i}_{\tilde n, n} \  \frac{n_i }{N_c + k}  
\end{split}
\label{eq:Kexpression}
\end{equation}
for $\tilde n$ that satisfies $\sum_i \tilde n_i = N_c$, and
$\mathcal K_{\tilde n,n} = 0$ otherwise.

\subsubsection{Total Transition, $\mathcal K \mathcal C \mathcal T$ }

We write down the matrix describing the total transition of the population dynamics (see Eq.~(\ref{eq:time_Evolution})). From the obtained expressions of $\mathcal K$, $\mathcal C$, $\mathcal T$, we calculate $\mathcal K \mathcal C \mathcal T$
%, which becomes
\begin{equation}
\begin{split}
&  (\mathcal K \mathcal C \mathcal T)_{\tilde n, n} = \delta_{\tilde n, n} \\
&   +  dt    \sum_{i=0}^{M-1} n_i  \sum_{j=0, (j\neq i)}^{M-1} \left [ w(i \rightarrow  j) + s\, \tilde w_{n}(i \rightarrow  j) \right ] \\ 
& \qquad \qquad \times \left [  \delta _{\tilde n_i,n_i-1}   \delta _{\tilde n_j,n_j+1} \ \delta ^{i,j}_{\tilde n,n} \  -  \delta _{\tilde n,n}    \right ],
\end{split}
\label{eq:KCTexpression}
\end{equation}
where the population-dependent transition rate $\tilde w_n(i\rightarrow j)$ is given as
\begin{equation}
\tilde w_n (i \rightarrow  j) = \frac{n_j  }{N_c} \left [ \alpha_j  \delta_{j\in \Omega^{(+)}} \frac{N_c}{N_c + 1} - \alpha_i \delta_{i\in \Omega^{(-)}}  \frac{N_c}{N_c - 1} \right ].
\end{equation}
The comparison of the expression (\ref{eq:KCTexpression}) with the original part $\mathcal T$ provides an insight into the obtained result. The jump ratio
$w(i\rightarrow j)$ in the original dynamics is replaced by $w(i\rightarrow j) + s\, \tilde w_n(i\rightarrow j)$ in the population dynamics algorithm. We note that this transition rate depends on the population $n$, meaning that we cannot get a closed equation for this modified dynamics at the level of the states $i$ in general.
We finally remark that the transition matrix $\sigma(n\rightarrow \tilde n)$ for the continuous-time limit is directly derived from (\ref{eq:KCTexpression}) as
\begin{equation}
\begin{split}
& \sigma(n\rightarrow \tilde n)  =   \sum_{i=0}^{M-1} n_i  \sum_{j=0, (j\neq i)}^{M-1} \left [ w(i \rightarrow  j) + s \tilde w_{n}(i \rightarrow  j) \right ] \\ 
& \qquad \qquad \qquad  \times \left [  \delta _{\tilde n_i,n_i-1}   \delta _{\tilde n_j,n_j+1} \ \delta ^{i,j}_{\tilde n,n}    \right ].
\label{eq:sigma}
\end{split}
\end{equation}

\subsection{Derivation of the Large Deviation Results \\ in the $N_c\rightarrow \infty$ Asymptotics}
\label{Subsec:DerivationLargeDeviationEstimator}

In this subsection, we study the $N_c \rightarrow \infty$ limit for the transition matrix of rates $\sigma(n\rightarrow \tilde n)$, and derive the validity of the population dynamics algorithm.

\subsubsection{The Estimator of the Large Deviation Function}
\label{subsubsec:estimator}
One of the ideal implementations of the population dynamics algorithm is as follows: We make copies of each realization (clone) at the end of simulation, where the number of copies for each realization is equal to the exponential weight $e^{-s\tau B(\tau)}$ in Eq.~(\ref{eq:defpsi}) (so that we can discuss an ensemble with this exponential weight without multiplying the probability by it). 
In this implementation, the number of clones grows (or decays) exponentially proportionally as $\left \langle e^{-s\tau B(\tau)} \right \rangle$ by definition. In real implementations of the algorithm, however, since taking care of an exponentially large or small number of clones  can cause numerical problems, one rather keeps the total number of clones to a constant $N_c$ at every time step, as seen in (\ref{eq:nufac}). 
Within this implementation, we reconstruct the exponential change of the total number of clones as follows:
%, which could have been directly observed if the total number of clones had not been kept constant. 
%a non-constant population algorithm would present. %in order to make a correspondence with $\left \langle e^{-s\tau B(\tau)} \right \rangle$, 
%
We compute the average of cloning ratio (see the beginning of Sec.~\ref{subsubsec:CloningRatio} for its definition) at each cloning step, and
we store the product of these ratios along the cloning steps. At final time, this product gives the empirical estimation of total (unnormalized) population during the whole duration of the simulation~\cite{1751-8121-49-20-205002_2016}, {\it i.e.}~an estimator of  $\left \langle e^{-s\tau B(\tau)} \right \rangle$. One thus estimates the CGF $\psi(s)$ given in Eq.~(\ref{eq:defpsi})~\cite{giardina_direct_2006,lecomte_numerical_2007,
JulienNaturephys_2007,tailleur_simulation_2009,giardina_simulating_2011,
1751-8121-46-25-254002_2013,1751-8121-49-20-205002_2016} as the logarithm of this reconstructed population, divided by the total time.

In our formulation, the average cloning ratio is given as $\sum _i n_i e^{- s dt b_i}/N_c$, and thus the multiplication over whole time interval reads $ \prod_{t=0}^{\tau/dt}    
 \{ n_i(t) e^{- s dt b_i}/N_c\}$.
Because we empirically assume that the CGF estimator converges to $\psi(s)$ in the $N_c, \tau \rightarrow \infty$ limit, 
%Therefore, the estimator of $\psi(s)$ in our formulation is given as the logarithm of this quantity divided by $\tau$, i.e., 
the following equality is expected to hold in probability 1:
\begin{equation}
\psi(s)  
\stackrel{?}{=}
 \lim_{N_c \rightarrow \infty} \lim_{\tau \rightarrow \infty} \frac{1}{\tau} \sum_{t=0}^{\tau/dt}  \ \log   
\sum _i \frac{n_i(t) e^{- s dt b_i}}{N_c}
 + O(dt). %V: i think this is not necessary. T: Yes, we need it, otherwise, the equality is not correct.
\label{eq:expected1}
\end{equation}
Since the dynamics of the population $n$ is described by a Markov process, ergodicity is satisfied, \emph{i.e.},~time averages can be replaced by the expected value with respect to the stationary distribution function. Applying this result to the right-hand side of (\ref{eq:expected1}), we obtain
\begin{equation}
\begin{split}
& \lim_{\tau \rightarrow \infty}  \frac{1}{\tau} \sum_{t=0}^{\tau/dt} 
 \ \log   
\sum _i \frac{n_i(t) e^{- s dt b_i}}{N_c}  \\
& = \frac{1}{dt}
\sum_{n}P_{n}^{\rm st} \ \log   
\sum _i \frac{n_i e^{- s dt b_i}}{N_c} 
+ \mathcal O(dt), % V not necesary isn't it? Yes, it is nessesary, because P_{n}^{\rm st} is the stationary distribution function in the dt->0 limit
\end{split}
\end{equation}
where $P_{n}^{\rm st}$ is the stationary distribution function of the population $n$ in the $dt \rightarrow 0$ limit, (namely, $P_{n}^{\rm st}$ is  the stationary distribution of the dynamics of transition rates $\sigma(n\rightarrow \tilde n)$). 
By expanding this right-hand side with respect to $dt$, we rewrite the expected equality (\ref{eq:expected1}) as
\begin{equation}
\psi(s)  \stackrel{?}{=} - s \lim_{N_c \rightarrow \infty} \sum_{n}P_{n}^{\rm st} \sum_{i}\frac{n_i b_i}{N_c} + O(dt).
\label{eq:expected2}
\end{equation}
where we used that $\sum_i n_i=N_c$ is a conserved quantity.
Below we demonstrate that this latter equality (\ref{eq:expected2}) is satisfied by analyzing the stationary distribution function $P_n^{\rm st}$.

\subsubsection{The Connection between the Distribution Functions \\ of the Population and of the Original System}

From the definition of the stationary distribution function $P_n^{\rm st}$, we have
\begin{equation}
\sum_{\tilde n} P_{\tilde n}^{\rm st} \sigma(\tilde n\rightarrow  n) - \sum_{\tilde n}P_n^{\rm st} \sigma(n\rightarrow \tilde n) = 0,
\label{eq:stationaryMaster}
\end{equation}
(which is a stationary Master equation.) 
In this equation, we use the explicit expression of $\sigma$ shown in (\ref{eq:sigma}). By
denoting by $n^{j \rightarrow i}$ the configuration where one clone in the state $j$ moves to the state $i$: $n^{j \rightarrow i} \equiv (n_0, n_1, \cdots, n_i + 1, \cdots, n_j -1, \cdots, n_{M-1})$, and the stationary master equation (\ref{eq:stationaryMaster}) is rewritten as
\begin{equation}
\sum_{i,j (i\neq j)}\left [  f_{i \rightarrow j}(n^{j\rightarrow i}) - f_{i \rightarrow j}(n) \right ] = 0,
\label{eq:derivation1}
\end{equation}
where we defined $f_{i \rightarrow j}(n)$ as
\begin{equation}
f_{i \rightarrow j}(n) = P_{n}^{\rm st} n_i \left [w(i \rightarrow j) + s \tilde w_n(i \rightarrow j) \right ].
\end{equation}
Now we multiply expression (\ref{eq:derivation1}) by $n_k$ ($k$ is arbitrary from $k=0,1,2,\cdots, M-1$), and sum it over all configurations $n$:
\begin{equation}
\sum_{n}\sum_{i,j (i\neq j)} n_k \left [  f_{i \rightarrow j}(n^{j\rightarrow i}) - f_{i \rightarrow j}(n) \right ] = 0.
\label{eq:derivation2}
\end{equation}
We can change the dummy summation variable $n$ in the first term to $n^{i\rightarrow j}$, which leads to
$\sum_{n}\sum_{i,j (i\neq j)} (n^{i\rightarrow j})_k    f_{i \rightarrow j}(n) $. Since the second term has almost the same expression as the first one except for the factor $n_k$, the sum in (\ref{eq:derivation2}) over the indices $(i,j)$, where none of $i$ nor $j$ is equal to $k$, becomes 0. The remaining term in (\ref{eq:derivation2}) is thus
\begin{equation}
\begin{split}
0 & = \sum_{n} \sum_{j(j\neq k)} \left ( (n^{k\rightarrow j})_k -  n_k\right ) f_{k \rightarrow j}(n)  \\
& + \sum_{n} \sum_{i(i\neq k)} \left ( (n^{i\rightarrow k})_k -  n_k\right ) f_{i \rightarrow k}(n).
\end{split}
\end{equation}
Using the definition of $n^{i \rightarrow j}$ in this equation, we arrive at
\begin{equation}
0  = \sum_{n} \bigg [  \sum_{i(i\neq k)}  f_{i \rightarrow k}(n) - f_{k \rightarrow i}(n)  \bigg].
\label{eq:derivation3}
\end{equation}
This equation (\ref{eq:derivation3}) connects the stationary property of the population dynamics (described by the occupation states $n$) and the one in the original system (described by the states $i$). 

The easiest case where we can see this connection is when $s=0$. By defining the empirical occupation probability of the original system as $p_i \equiv \sum_{n}P_{n}^{\rm st}n_i/N_c$, Eq.~(\ref{eq:derivation3}) leads to the following (stationary) master equation for $w(i\rightarrow j)$:
\begin{equation}
0 = \sum _{j}p_j w(j\rightarrow i)  - \sum_{j}p_i w(i\rightarrow j)  \quad\textnormal{(for $s=0$)}
\label{eq:mastereqpiemp}
\end{equation}
This is valid for any $N_c$, meaning that, for original Monte-Carlo simulations in $s=0$, the empirical probability~$p_i$ is exactly equal to the steady-state probability, as being the unique solution of~\eqref{eq:mastereqpiemp}.
It means that there are no systematic errors in the evaluation of $p_i $ (see the introduction of this paper for the definition of the term ``systematic errors'').  However, in the generic case $s\neq 0$, this property is not satisfied. One thus needs to understand the $N_c \rightarrow \infty$ limit to connect the population dynamics result with the large deviation property of the original system.

\subsubsection{Justification of the Convergence of the Large Deviation Estimator as Population Size becomes Large}
In order to take the $N_c \rightarrow \infty$ limit, we define a scaled variable $x_i$ as $n_i/N_c$. With keeping this occupation fractions $x_i$ to be $\mathcal O(1)$, 
we take the $N_c \rightarrow \infty$ limit 
in (\ref{eq:derivation3}), which leads to 
\begin{equation}
\begin{split}
0 =  \sum_n P_n^{\rm st} & \Bigg [ 
 \sum _{j}x_j w(j\rightarrow i)  - \sum_{j}x_i w(i\rightarrow j)  \\
 & - s ~ x_i \left ( b_i  -  \sum_k x_k b_ k  \right )    \Bigg ]
 +  \mathcal O(1/N_c).
\end{split}
\label{eq:derivation4}
\end{equation}
Inspired by this expression, we define a matrix $L_{i,j}^s$ as
\begin{equation}
L_{i,j}^s = w(j\rightarrow i) - \delta _{i,j} \left (\sum_{k} w(i \rightarrow k)  + s ~ b_i\right ),
\end{equation}
and a correlation function between $x_i$ and $x_j$ as
\begin{equation}
c_{i,j} = \sum_{n}x_ix_j P_{n}^{\rm st} - p_i p_j,
\end{equation}
(where we recall $p_i \equiv \sum_n x_i P_{n}^{\rm st}$). 
From these definitions, (\ref{eq:derivation4}) is rewritten as
\begin{equation}
\sum_{j}p_j L_{i,j}^{s}  = - s p_i \sum_{k}  p_kb_k - s \sum_{k}c_{i,k} b_k  + \mathcal O \left ( \frac{1}{N_c} \right ).
\label{eq:generaleigenvalue}
\end{equation}
Since $x_i$ is an averaged quantity (an arithmetic mean) with respect to the total number of clones ($x_i\equiv n_i/N_c$), we can safely assume that the correlation $c_{i,j}$ becomes 0 in $N_c \rightarrow \infty$ limit:
\begin{equation}
\lim_{N_c\rightarrow \infty}c_{i,k}=0.
\label{eq:Assumption}
\end{equation}
 (For more detailed discussion of why this is valid, see the description after Eq.~(\ref{eq:systemsizeexpansion})).
Thus, by defining $p_i^{\infty}\equiv \lim_{N_c \rightarrow \infty} p_i$, we obtain
\begin{equation}
\sum_{j}p_j^{\infty} L_{i,j}^{s}  = - s p^{\infty}_i \sum_{k}  p^{\infty}_kb_k.
\end{equation}
From the Perron-Frobenius theory, the positive eigenvector of the matrix $L_{i,j}^{s}$ is unique  and corresponds to its eigenvector of largest eigenvalue (in real part). This means that $-s \sum_{k}p_{k}^{\infty } b_k$ is the largest eigenvalue of the matrix $L_{i,j}^{s}$.  Finally, by recalling that
the largest eigenvalue of this matrix $L_{i,j}^{s}$ is equal to the generating function $\psi(s)$ (see Ref.~\cite{garrahan_first-order_2009} for example), we have finally justified that the CGF estimator (\ref{eq:expected2}) is valid in the large-$N_c$ limit.

\subsection{Systematic Errors due to Finite $N_c$; Convergence Speed of the Large Deviation Estimator as $N_c\to\infty$}
\label{subsec:systemsizeexpansion}

In the introduction of this paper, 
we defined the systematic errors as the deviations of the large deviation estimator from the correct value due to a finite number of clones~$N_c$. 
From (\ref{eq:expected2}), we quantitatively define this systematic error $\epsilon_{\rm sys} $ as
\begin{equation}
\epsilon_{\rm sys} \equiv \left |  \psi(s) + s  \sum_i p_i b_i \right |.
\label{eq:systematicdef}
\end{equation}
From a simple argument based on a system size expansion, we below show that this $\epsilon_{\rm sys}$ is of order $\mathcal O(1/N_c)$.

We first show that one can perform a system size expansion (as, \emph{e.g.} in~van Kampen~\cite{tagkey2007iii}) for the population dynamics. In (\ref{eq:derivation1}), by recalling the definition of the vector $x$ as $x=n/N_c$, and by denoting $\tilde P^{\rm st}(x) = P^{\rm st}_{x N_c} $, we obtain
\begin{equation}
\begin{split}
0 = & \sum_{i,j (i\neq j)}\sum_{r=1}^{\infty}\frac{1}{r !} \frac{1}{N_{c}^r} \left (\frac{ \partial}{\partial x_i} - \frac{\partial}{\partial x_j} \right )^r x_i  \tilde P^{\rm st}(x) \\
 & \qquad \qquad \quad \times \left [ w(i\rightarrow j) + s \tilde w_n(i\rightarrow j)|_{n=xN_c}  \right ].
\label{eq:systemsizeexpansion}
\end{split}
\end{equation}
This indicates that the stochastic process governing the evolution of $x$ becomes deterministic in the $N_c \rightarrow \infty$ limit.  The deterministic trajectory for $x$ is governed by a differential equation derived from the sole term  $r=1$ in the expansion (\ref{eq:systemsizeexpansion}) (see \emph{e.g.}~Sec.~3.5.3 {\it Deterministic processes - Liouville's Equation} in Ref.~\cite{opac-b1079113} for the detail of how to derive this property). Thus if $x$  converges to a fixed point as $N_c$ increases, {\it which is normally observed in implementations of cloning algorithms}, the assumption $(\ref{eq:Assumption})$ is satisfied.

From the expression of $\epsilon_{\rm sys}$, we see that the dependence in~$N_c$ comes solely from $p_i$, which can be calculated from the first order correction of $P^{\rm st}_n$ (at large $N_c$). The equation to determine $P^{\rm st}_n$ is the stationary master equation (\ref{eq:stationaryMaster}) or equivalently, the system-size expansion formula (\ref{eq:systemsizeexpansion}). We expand the  jump ratio  $w(i\rightarrow j) + s \tilde w_n(i\rightarrow j)$ in (\ref{eq:systemsizeexpansion}) with respect to $1/N_c$ as:
\begin{equation}
\begin{split}
& w(i\rightarrow j) + s \tilde w_n(i\rightarrow j)  \\
& = w(i\rightarrow j) + s \tilde w_x^\infty (i \rightarrow j) + \frac{s}{N_c} \delta w_x(i\rightarrow j) + \mathcal O(1/N_c^2),
\label{eq:wexpansion}
\end{split}
\end{equation}
where $\tilde w_x^\infty (i \rightarrow j)$ and $\delta w_x (i \rightarrow j)$ are defined as
\begin{equation}
\tilde w_x^\infty (i \rightarrow j)  = x_j \left [\alpha_j \delta_{j\in \Omega^{(+)}} - \alpha_i \delta_{i\in \Omega^{(-)}}  \right ]
\end{equation}
and
\begin{equation}
\delta w_x (i \rightarrow j)  = - x_j \left [\alpha_j \delta_{j\in \Omega^{(+)}} + \alpha_i \delta_{i\in \Omega^{(-)}}  \right ].
\end{equation}
By substituting (\ref{eq:wexpansion}) into the system-size expansion formula (\ref{eq:systemsizeexpansion}) and performing a perturbation expansion, we find that a first-order correction of $p$ is naturally of order $\mathcal O(1/N_c)$, \emph{i.e.}~$\epsilon_{\rm sys}=\mathcal O(1/N_c)$. For a practical scheme of how to implement this perturbation on a specific example, see Sec.~\ref{subsection:systematicerrors_twostate}.
In %Sec.~IV of 
our companion paper~\cite{SecondPart}, the scaling analysis of the $1/N_c$ correction is shown to hold numerically with the continuous-time cloning algorithm (see Sec.~\ref{subsubsec:differenceContinuousTime}). We also show that the $1/N_c$ correction behavior remains in fact valid at {\it finite time}~\cite{SecondPart}, an open question that remains to be investigated analytically.

\begin{table}
\begin{center}
          \caption{\label{Table:Numericalerrors} Magnitudes of the numerical errors}
          \begin{tabular}{c||c}
                     	  			& Magnitude of errors				 \\  \hline \hline
    Systematic errors	 	 & 		$\mathcal O(1/N_c)$   \\  \hline
  Numerical errors			&              $\mathcal O(1/(\tau N_c))$  \\ \hline
          \end{tabular}  
        \end{center}
 \end{table}

\subsection{Remarks}

Here, we discuss some remarks on the formulation presented in this section. 

\subsubsection{Relaxing the Condition $dt=\Delta t$}
\label{subsubsec:dtDeltat}
In Sec.~\ref{subsubsec:CloningRatio}, we set the discretization time of the process $dt$ to be equal to the time interval for cloning $\Delta t$, and we took the $dt=\Delta t \rightarrow 0$ limit at the end. We note that the condition $\Delta t = dt $ is not necessary if both limits $\Delta t \rightarrow 0$ and $d t \rightarrow 0$ (with $dt<\Delta t$) are taken at the end. 
This is practically important, because we can use the continuous-time {\it process} to perform the algorithm presented here by setting $dt=0$ first, and $\Delta t \rightarrow 0$ limit afterwards. More precisely, replacing $dt$ by $\Delta t$ in the matrix $\mathcal C$ and $\mathcal K$, we build a new matrix $\mathcal K \mathcal C (\mathcal T^{\Delta t/dt})$.
Taking the $dt \rightarrow 0$ limit in this matrix while keeping $\Delta t$ non-infinitesimal (but small), this matrix represents the population dynamics algorithm of a continuous-time process with a finite cloning time interval $\Delta t$. The arguments presented in this section can  then be applied in the same way, replacing $dt$ by $\Delta t$.
We note that the deviation due to a non-infinitesimal $\Delta t$ should thus appear as $O(\Delta t)$ [see Eq.~(\ref{eq:expected1})  for example].

\subsubsection{A Continuous-time Algorithm Used in the companion paper}
\label{subsubsec:differenceContinuousTime}
The $\Delta t \rightarrow 0$ limit is the key point in the formulation developed in this section.
Thanks to this limit, upon each cloning step, the total number of clones $\sum_{j=0}^{M-1}n_j$ always varies only by $\pm 1$, which makes the expression of the matrices $\mathcal C$ and $\mathcal K$ simple enough to develop the arguments presented in Secs.~\ref{Subsec:DerivationLargeDeviationEstimator} and~\ref{subsec:systemsizeexpansion}.
%understand the resulting scalings (see (\ref{eq:Cdt0limit}) and (\ref{eq:Kexpression})). 
Furthermore, during the time interval $\Delta t$ separating  two cloning steps, the configuration is changing {\it at most once}. 
The process between cloning steps is thus simple, which allows us to represent the corresponding time-evolution matrix as $\mathcal T$ (by replacing $dt$ by $\Delta t$ as explained in Sec.~\ref{subsubsec:dtDeltat} above). 
Generalizing our analytical study to a cloning dynamics in which the limit $\Delta t\to 0$ is not taken is therefore a very challenging task, which is out of the scope of this paper.

However, interestingly, in the companion to this paper~\cite{SecondPart} we observe numerically that our predictions for the finite-time and finite-population scalings are still valid in a different version of algorithm
for which $\sum_{j=0}^{M-1}n_j$ can vary by an arbitrary amount --~supporting the hypothesis that the analytical arguments that we present here could be extended to more general algorithms.
More precisely, we use a continuous-time version of the algorithm~\cite{lecomte_numerical_2007} %(see Sec.~II of Part~II for the detail of the algorithm), 
to study numerically an observable of `type~A'~\cite{garrahan_first-order_2009}.
This version of the algorithm differs from that considered in this paper, in the sense that 
the cloning steps are separated by {\it non-fixed} non-infinitesimal time intervals. These time intervals are distributed exponentially, in contrast to the fixed ones taken in here (where $\Delta t$ is a constant).
%after each selection step, a copy in the population can have strictly more than one offspring. This is because 
%
This results in an important difference: The effective interaction between copies due to the cloning/pruning procedure is unbounded (it can \emph{a~priori} affect any proportion of the population), while in the algorithm of the present paper, this effective interaction is restricted to a maximum of one cloning/pruning event in the $\Delta t \to 0$ limit.
We stress that the $dt\to 0$ limit of the cloning algorithm studied here with a fixed $\Delta t$ \emph{does not yield the continuous-time cloning algorithm}, stressing that these two versions of the population dynamics present essential differences.
%

%In summary, the analytical method that we present in this Part~I allows us to analyze only the fixed-time-$\Delta t$ version of the cloning algorithm in the $\Delta t \rightarrow 0$ limit. However, we observe numerically in Part~II that the same finite-time and finite-population size scalings are present in its continuous-time version (where $\Delta t$ is distributed exponentially), illustrating their universal character.

\section{Stochastic Errors: Large Deviations of the Population Dynamics}
\label{Sec:largedeviation_largedeviation}

In the previous section, we formulated the population dynamics algorithm as a birth-death process and evaluated the systematic errors (which are the deviation of the large deviation estimator from the correct value) due to a finite number of clones (Table~\ref{Table:Numericalerrors}). In this section, we focus on stochastic errors corresponding to the run-to-run fluctuations of the large deviation estimator within the algorithm, at fixed $N_c$ (see the introduction of this paper for the definition of the terms {\it stochastic errors} and {\it systematic errors}).

In order to study stochastic errors, we formulate the large deviation principle of the large deviation estimator. In the population dynamics algorithm, the CGF estimator to measure is the time-average of the average cloning ratio of the population (see Sec.~\ref{subsubsec:estimator}):
\begin{equation}
\psi_{N_c,\tau}(s) \equiv - s \frac{1}{\tau} \int_{0}^{\tau} dt \sum_{i=0}^{M-1}   \frac{n_i(t) b_i}{N_c}.
\label{eq:estimetor2}
\end{equation}
As $\tau$ increases, this quantity converges to the expected value (which depends on $N_c$) with probability 1. However whenever we consider a finite $\tau$, dynamical fluctuations are present, and there is a probability that this estimator deviates from its expected value. Since the population dynamics in the occupation states $n$ is described by a Markov process, the probability of these deviations are themselves described by a large deviation principle~\cite{Touchette20091,opac-b1093895}: By denoting by $\rm Prob (\psi)$ the probability of $\psi_{N_c,\tau}(s)$, one has:
\begin{equation}
{\rm Prob} (\psi) \sim \exp \left (  - \tau I_{N_c,s}(\psi) \right ),
\end{equation}
where $I_{N_c,s}(\psi)$ is a large deviation ``rate function'' (of the large deviation estimator). 
To study these large deviations, we can apply a standard technique using a biased evolution operator for our population dynamics: For a given Markov system, to calculate large deviations of additive quantities such as~\eqref{eq:estimetor2}, one biases the time-evolution matrix with an exponential factor~\cite{opac-b1093895}.  
Specifically, by defining the following matrix
\begin{equation}
L^{h}_{\tilde n,n} = \sigma(n \rightarrow \tilde n) - \delta_{\tilde n, n} \sum_{n^{\prime}}\sigma(n \rightarrow n^{\prime}) -  h s \sum_{i=0}^{M-1}   \frac{n_i b_i}{N_c}.
\label{eq:DefinitionLh}
\end{equation}
and by denoting the largest eigenvalue of this matrix $G(h,s)$ (corresponding, as a function of $h$, to a scaled cumulant generating function for the observable~\eqref{eq:estimetor2}), the large deviation function $I_{N_c,s}(\psi)$ is obtained as the Legendre transform $\sup_{h} \left [h \psi -  G(h,s) \right ]$.
In the companion paper~\cite{SecondPart}, we show that a quadratic approximation of the rate function~$I_{N_c,s}(\psi)$ (\emph{i.e.}~a Gaussian approximation) can be estimated directly from the cloning algorithm.
% in the region where fluctuations of the CGF estimator are approximately Gaussian.

We consider the scaling properties of $I_{N_c,s}$ in the large-$N_c$ limit. For this, we define a scaled variable $\tilde h \equiv h /N_c$ and a scaled function $\tilde G(\tilde h,s) \equiv G(\tilde h N_c,s)/N_c$. If this scaled function $\tilde G(\tilde h,s) \equiv G(\tilde h N_c,s)/N_c$ is well-defined in the $N_c \rightarrow \infty$ limit (which is natural as checked in the next paragraph), then we can derive that $I_{N_c,s}$ has the following scaling:
\begin{equation}
I_{N_c,s}(\psi) = N_c I_{s}(\psi) + o(N_c)
\label{eq:scalingNcINcs}
\end{equation}
or equivalently,
\begin{equation}
{\rm Prob}(\psi) \sim e^{-\tau N_c  I_{s}(\psi)},
\label{eq:largedeviation_largedeviation}
\end{equation}
where $I_{s}(\psi) = \max_{\tilde h} \left [ \tilde h \psi -  \tilde G(\tilde h,s)   \right ]$.
The scaling form~\eqref{eq:scalingNcINcs} is validated numerically in Ref.~\cite{SecondPart}.
From this large deviation principle, we can see that the stochastic errors of the large deviation estimator is of $\mathcal O(1/(N_c \tau))$ as shown in Table~\ref{Table:Numericalerrors}.

In the largest eigenvalue problem for the transition matrix (\ref{eq:DefinitionLh}), by performing a system size expansion (see Sec.~\ref{subsec:systemsizeexpansion}), we obtain
\begin{equation}
\begin{split}
\tilde G(\tilde h,s) = & \sum_{i,j (i\neq j)}  \left (\frac{ \partial}{\partial x_i} - \frac{\partial}{\partial x_j} \right ) x_i  q(x) \\
 & \qquad \qquad \times \left [ w(i\rightarrow j) + s \tilde w_x^{\infty}(i\rightarrow j)  \right ] \\
& - \frac{\tilde h}{s} s \sum_i x_i b_i q(x) + \mathcal O(1/N_c),
\label{eq:systemsizeexpansion2}
\end{split}
\end{equation}
where $q(x)$ is the right-eigenvector associated to the largest eigenvalue of $L_{\tilde n,n}^h$ (represented as a function of $x\equiv n/N_c$). 
The first order of the right-hand side is of order $\mathcal O(N_c^0)$, so that $\tilde G(\tilde h,s)$ is also of order $\mathcal O(N_c^0)$ in $N_c \rightarrow \infty$. (For an analytical example of the function $\tilde G(\tilde h, s)$, see Sec.~\ref{subsection:largedeviationsInPopTwoState}).

\section{Example: \\ A Simple Two-State Model}
\label{Section:Demonstrations}

In this section, to illustrate the formulation that we developed in the previous sections, we consider a simple two state model. 
In this system, the dimension of the state $i$ is two ($M=2$) and the transition rates $w( i \rightarrow j)$ are
\begin{equation}
w(0  \rightarrow 1) = c,
\end{equation}
\begin{equation}
w(1  \rightarrow 0) = d
\end{equation}
with positive parameters $c, d$  and $w(i  \rightarrow  i) = 0$.
In this model, the quantity
$\alpha_{i}$ defined in (\ref{def:eqalphai}) becomes
\begin{equation}
\alpha_{i} = \delta_{i,0} \frac{n_1}{N_c}\left (b_1 - b_0 \right ) +  \delta_{i,1} \frac{n_0}{N_c} (b_0 - b_1).
\end{equation}
Hereafter, we assume that $b_1>b_0$ without loss of generality. From this, the space $\Omega^{(\pm)}$ is determined as $\Omega^{(+)}=\{ 0 \}$ and $\Omega^{(-)}=\{ 1 \}$, which leads to the jump ratio $\tilde w_{n}(i \rightarrow j)$ as
\begin{equation}
\tilde w_{n}(i \rightarrow j) =\delta_{i,1}\delta_{j,0}  \frac{n_0}{N_c}(b_1 - b_0) \left [\frac{n_1}{N_c+1} + \frac{n_0}{N_c -1} \right ].
\end{equation}
Finally, from the conservation of the total population: $n_0+n_1=N_c$, we find that the state of the population $n$ can be
uniquely determined
 by specifying only the variable $n_0$.
Thus the transition rate for the population dynamics is a function of $n_0$ (and $\tilde n_0$), $\sigma (n_0 \rightarrow \tilde n_0)$, which is derived as
\begin{equation}
\begin{split}
& \sigma(n_0\rightarrow \tilde n_0) =   \delta_{\tilde n_0, n_0 + 1 } \bigg [ (N_c - n_0) d \\
&+ k(n_0,N_c - n_0) \Big( \frac{n_0}{N_c -1} + \frac{N_c - n_0}{N_c + 1} \Big) \bigg ] + \delta_{\tilde n_0, n_0 - 1 } \ n_0 \,c,
\end{split}
\end{equation}
where we have defined
\begin{equation}
k(n_0,n_1) = \frac{n_0 n_1}{N_c} s \left [b_1 - b_0 \right ].
\end{equation}

\subsection{Systematic Errors}
\label{subsection:systematicerrors_twostate}
We first evaluate the systematic errors (see Sec.~\ref{subsec:systemsizeexpansion}). For this, we consider the distribution function $P_{n}^{\rm st}$. 
Since the system is described by a one dimensional variable $n_0$ restricted to $0\leq n_0 \leq N_c$, the transition rates $\sigma(n_0 \rightarrow \tilde n_0)$ satisfy the detailed balance condition:
\begin{equation}
P_{n_0}^{\rm st}  \sigma(n_0 \rightarrow n_0 + 1)  = P_{n_0+1}^{\rm st} \sigma(n_0 + 1 \rightarrow n_0).
\end{equation}
We can solve this equation exactly, but to illustrate the large-$N_c$ limit, it is in fact sufficient to study the solution in an expansion $1/N_c\ll 1$. The result is
\begin{equation}
P^{\rm st}_{x N_c} = C \exp \left [ - N_c I_{\rm conf}(x) + \delta I(x) + \mathcal O(1/N_c)  \right ]
\end{equation}
(with here $x\equiv n_0 / N_c$), where, explicitly
\begin{equation}
\begin{split}
 I_{\rm conf}(x) = & x + \log(1-x) - \frac{d \log \left [d+(b_1-b_0) s x\right ]}{(b_1-b_0) s} \\
&- x \log \left [  \frac{1}{cx}(1-x)\left (d + (b_1 - b_0) s x \right ) \right ]
\end{split}
\end{equation}
and
\begin{equation}
\begin{split}
\delta I(x)  & =  - x  - \frac{2 d x}{(b_1-b_0) s} + x^2 - \log x  \\
& + \frac{2 d^2 \log \left [ d + (b_1 - b_0)s x\right ] }{(b_1-b_0)^2 s^2} + \frac{d \log \left[ d + (b_1-b_0) s x \right ]}{(b_1-b_0)s}.
\end{split}
\end{equation}

We now determine the value of $x$ that minimizes $ - N_c I_s(x) + \delta I(x)$, which leads to a finite-size correction (\emph{i.e.}~the systematic errors) of the population dynamics estimator. Indeed, denoting this optimal value of $x$ by $x_{N_c}^*$, the large deviation estimator is obtained as
\begin{equation}
\psi_{N_c}(s) = - s \left [ x^*_{N_c} b_0 +(1-x^*_{N_c})b_1  \right ]
\end{equation}
(see Sec.~\ref{subsubsec:estimator}).
From a straightforward calculation based on the expressions $I_{\rm conf}(x)$ and $\delta I(x)$, we obtain
the expression of $x_{N_c}^*$ as
\begin{equation}
x^*_{N_c} = x^{*} + \frac{1}{N_c} \delta x^* + \mathcal O((1/N_c)^2),
\end{equation}
with
\begin{equation}
\begin{split}
x^{*} & = \frac{-c - d + (b_1-b_0) s }{2 (b_1 - b_0) s} \\
& + \frac{ \sqrt{4 d (b_1-b_0) s + \left [-c-d  + (b_1-b_0)s \right ]^2}}{2 (b_1 - b_0) s} 
\end{split}
\label{eq:xstar}
\end{equation}
and
\begin{equation}
\begin{split}
\delta x ^*  = &  \left (2 d + 2 (b_1 - b_0) s x^* \right )^{-1} \\
 & \times    \frac{2 c \left [ - d - (b_1 - b_0) s x^*  \left ( 1 +  x^*  - 2 (x^*)^2 \right ) \right ]}{\sqrt{4 d (b_1-b_0)s + [c+d-(b_1-b_0)s ]^2}  }.
\end{split}
\end{equation}
We thus arrive at
\begin{equation}
\begin{split}
 \psi(s)  = &     \frac{-c - d - (b_1 + b_0) s }{2 } \\
& +   \frac{ \sqrt{4 d (b_1-b_0) s + \left [-c-d  + (b_1-b_0)s \right ]^2}}{2 } 
\label{eq:PsiInfinitetime_InfiniteCopy}
\end{split}
\end{equation}
and
\begin{equation}
\begin{split}
\epsilon _{\rm sys} = & \frac{1}{N_c}   \frac{1}{\left | d+(b_1-b_0)sx^*\right |}  \\
& \times \left |  \frac{s c(b_0-b_1) \left ( d + (b_0 - b_1) s (x^* - 1)  x^* (1+2x^*) \right ) }{ \sqrt{4(b_1-b_0)d s + [c+d +(b_0 - b_1)s]^2}} \right |
\end{split}
\end{equation}
(see Eq.~(\ref{eq:systematicdef}) for the definition of the systematic error $\epsilon_{\rm sys}$.)
We check easily that the expression of $\psi(s)$ is the same as the one obtained from a standard method by solving the largest eigenvalue problem of a biased time-evolution operator (see for example, Ref.~\cite{1751-8121-49-20-205002_2016}).

\subsection{Stochastic Errors}
\label{subsection:largedeviationsInPopTwoState}

We now turn our attention to the stochastic errors. The scaled cumulant generating function $N_c \tilde G(\tilde h, s)$ is the largest eigenvalue of a matrix $L_{\tilde n,n}^{h}$ (see Eq.~(\ref{eq:DefinitionLh}) and the explanations around it). We then recall 
a formula to calculate this largest eigenvalue problem from the following variational principle:
\begin{equation}
\begin{split}
&\tilde G(\tilde h,s) \\
& = \sup_{\phi>0} \sum_{n}  p_{\rm st}(n_0)\phi(n_0)^2  \Bigg [ \frac{\sigma(n\rightarrow n+1)}{N_c} \left ( \frac{\phi(n_0 + 1)}{\phi(n_0)} - 1 \right ) \\
& + \frac{\sigma(n\rightarrow n-1)}{N_c } \left (   \frac{\phi(n_0 - 1)}{\phi(n_0)} - 1 \right ) - s \tilde h \frac{\sum_{i}n_i b_i}{N_c^2}  \Bigg ].
\end{split}
\end{equation}
(See, \emph{e.g.},  Appendix G of Ref.~\cite{PhysRevE.84.061113} or Ref.~\cite{garrahan_first-order_2009} for the derivation of this variational principle). By following the usual route to solve such equations (see, \emph{e.g.}, Sec.~2.5 of Ref.~\cite{1742-5468-2014-10-P10001}), we obtain
\begin{equation}
\begin{split}
\tilde G(\tilde h,s) =   \sup_{x} & \Bigg [ - \left ( \sqrt{(1-x)(d+(b_1 - b_0)s x)} - \sqrt{c x} \right )^2 \\
 & - s \tilde h \left [  x b_0  + (1-x) b_1 \right ]    \Bigg ].
\end{split}
\end{equation}
Thus, $\tilde G(\tilde h,s)$ is well-defined, demonstrating that the large deviation principle (\ref{eq:largedeviation_largedeviation}) is satisfied. 
Furthermore, by expanding this variational principle with respect to $\tilde h $, we obtain
\begin{equation}
\tilde G(\tilde h,s) =\psi (s)  \tilde h  + \frac{\kappa_s}{2} \tilde h ^2  + \mathcal O(\tilde h^3),
\label{eq:expansionGtilde}
\end{equation}
where $ \psi (s)$ is given in (\ref{eq:PsiInfinitetime_InfiniteCopy}), and the variance $\kappa_s$ is given as
\begin{equation}
\begin{split}
\kappa_s = & c  +  \frac{c s(b_1 - b_0)}{\sqrt{4(b_1-b_0) s d + (c+d+(b_0 - b_1)s)^2}} \\
&  -  \frac{ c(c+d)^2 + c (b_0 - b_1)(c-3d)s }{c^2 + 2 c \left [ d + (b_0 - b_1) s \right ] + (d+(b_1-b_0)s)^2}.
\end{split}
\end{equation}
We note that the expansion (\ref{eq:expansionGtilde}) is equivalent to the following expansion of the large deviation function
 $I_{s}(\psi)$ (see (\ref{eq:largedeviation_largedeviation})) around the expected value $\psi (s)$:
\begin{equation}
I_{s}(\psi)  = \frac{(\psi - \psi(s))^2}{2\kappa_s}  + \mathcal O ((\psi - \psi_s)^3).
\end{equation}
The variance of the obtained large deviation estimator is thus $\kappa_s/(N_c \tau)$.

\subsection{A Different Large Deviation Estimator}
\label{subsec:Different large deviation estimator}

As an application of these exact expressions, we expand the systematic error $\epsilon _{\rm sys}$ and the stochastic error (variance) $\kappa_s$ with respect to $s$. 
A straightforward calculation leads to
\begin{equation}
\begin{split}
\epsilon _{\rm sys} N_c = &\Bigg |   \frac{2c(b_0 - b_1)}{c+d} s \Bigg | + \mathcal O(s^2) 
\label{eq:epsilon_expansion}
\end{split}
\end{equation}
and 
\begin{equation}
\kappa_s = \frac{2 (b_0 - b_1)^2 c d }{(c+d)^3} s^2 + \mathcal O(s^3).
\label{eq:kappa_expansion}
\end{equation}
We thus find that the first-order of the error $\epsilon _{\rm sys}$ scales as $\mathcal O(s)$ at small $s$, but that the variance $\kappa_s$ is of order $\mathcal O(s^2)$. From this scaling, as we explain below,  
one can argue that 
the following large deviation estimator can be better than the standard one for small $s$:
\begin{equation}
\tilde \Psi(s) \equiv   \frac{1}{\tau}   \log  \overline{ \prod_{t=0}^{\tau / dt}  
\sum _i \frac{n_i(t) e^{- s dt b_i}}{N_c} },
\label{Definition_Psi_s}
\end{equation}
where the overline represents the averaging with respect to the realizations of the algorithm. (Normally, this realization-average is taken \emph{after} calculating the logarithm, which corresponds to the estimator (\ref{eq:expected1}).) 
Mathematically, this average (\eqref{Definition_Psi_s}, before taking the logarithm) corresponds to a bias of the time-evolution matrix $\sigma$ as seen in (\ref{eq:DefinitionLh}) for $h=1$. This means that, in the limit $\tau \rightarrow \infty$ with a sufficiently large number of realizations, this averaged value behaves as $\tilde \Psi(s) \sim e^{\tau G(1,s)}$. By combining this result with the expansion (\ref{eq:expansionGtilde}), we thus obtain
\begin{equation}
\begin{split}
\lim_{\tau \rightarrow \infty} \lim_{\substack{ \text{many} \\ \text{realizations}}}
\tilde \Psi(s)
= \psi(s) + \frac{\kappa_s}{2} N_c^{-1} + \mathcal O(N_c^{-2})
\label{eq:G(1s)}
\end{split}
\end{equation}
(recalling $\tilde G = G/N_c$ and $\tilde h = h/N_c$).
When we consider small $s$, by recalling $\epsilon_{\rm sys}N_c=\mathcal O(s)$ and $\kappa_s=\mathcal O(s^2)$, we thus find that the deviations from the correct value are smaller in the estimator $\tilde \Psi(s)$ than in the normal estimator given in (\ref{eq:expected1}), which comes as a surprise because in~\eqref{Definition_Psi_s} the average and the logarithm are inverted with respect to a natural definition of the CGF estimator.

To use this estimator, we need to discuss the two following points. First, since the scaled cumulant generating function $G(1,s)$ has small fluctuations, one needs a very large number of realizations in order to attain the equality (\ref{eq:G(1s)}). The difficulty of this measurement is the same level as the one of direct observations of a large deviation function, see for example Ref.~\cite{PhysRevE.92.052104}. However, we stress that this point may not be fatal in this estimator, because we do not need to attain completely this equality, \emph{i.e.}~our aim is the zero-th order coefficient, $\psi(s)$, in (\ref{eq:G(1s)}). Second, we have not proved yet the scaling properties with respect to $s$, which are $\epsilon_{\rm sys}N_c=\mathcal O(s)$ and $\kappa_s=\mathcal O(s^2)$, in a general set-up aside from this simple two state model.  
We show in practice in Ref.~\cite{SecondPart} that for small values of~$s$, the estimator~\eqref{Definition_Psi_s} is affected by smaller systematic errors, in the numerical study of the  creation-annihilation process studied in this section.
We will focus on the generality of our observations on these points in a future study.

\section{Discussion}
\label{sec:discussion}

In this paper, we formulated a birth-death process that describes population dynamics algorithms and aims at evaluating numerically large deviation functions. We derived that this birth-death process leads generically to the correct large deviation results in the large limit of the number of clones $N_c \rightarrow \infty$. From this formulation, we also derived that the deviation of large deviation estimator from the desired value (which we called systematic errors) is small and proportional with $\mathcal O(N_c^{-1})$. Below, based on this observation, we propose a simple interpolation technique to improve the numerical estimation of large deviation functions in practical uses of the algorithm.

\subsection{An Interpolation Technique using the $\mathcal O(1/N_c)$ Scaling of the Systematic Error}
\label{subsec:Interporationtechnique}

Imagine that we now apply the population dynamics algorithm to a given system. We need to carefully consider the asymptotic limit of the two large parameters $\tau$ and $N_c$ in the convergence of the large deviation estimator (\ref{eq:estimetor2}). Indeed, what one needs to do in this simulation is, (\emph{i}) take the large-$\tau$ limit for a fixed $N_c$ and estimating the $\tau \rightarrow \infty$ value of the estimator for this fixed $N_c$, and then (\emph{ii}) estimate this large-$\tau$ value for several (and increasing) $N_c$, and finally estimate large-$\tau$-$N_c$ limit value. This is different from standard Monte-Carlo simulations, where one needs to consider only the large-$\tau$ limit, thanks to  ergodicity. 

Any method that can make the LDF estimation easier thus will be appreciated. 
Based on our observations, we know that the second part [(\emph{ii}) above] converges with an error proportional to $1/N_c$. 
Also, from the large deviation estimator  (\ref{eq:estimetor2}), one can easily see that the convergence speed with respect to $\tau$ for a fixed $N_c$ is  proportional to $1/\tau$ (\emph{i.e.}~the first part [(\emph{i}) above] converges proportionally to $1/\tau$). By using these $1/\tau$- and $1/N_c$-scalings for (\emph{i}) and (\emph{ii}), one can interpolate the large-$\tau$ and large-$N_c$ asymptotic value of the LDF estimator from the measured values for finite $\tau$ and $N_c$. 
We introduce this numerical method in practice in the companion paper~\cite{SecondPart}.
%, in Sec.~V.
%
We demonstrate numerically that the interpolation technique is very efficient in practice, by a direct comparison of the resulting estimation of the CGF to its analytical value, which is also available in the studied system.
We also stress that it is developed for a different cloning algorithm by using a continuous-time population dynamics~\cite{lecomte_numerical_2007}
(see Sec.~\ref{subsubsec:differenceContinuousTime} for the description of the conceptual difference).
From these results, we conjecture that the validity of the large-$\tau$ and large-$N_c$ scalings is very general and independent of the details of the algorithm.

\bigskip

\subsection{Open Questions}
\label{subsec:openquestions}

We mention two open questions. The first question is about the precise estimate of the error due to a non-infinitesimal time interval $\Delta t$ between cloning steps:
As explained in Secs.~\ref{subsubsec:dtDeltat} and \ref{subsubsec:differenceContinuousTime}, 
taking the $\Delta t \rightarrow 0$ limit is important in our analysis, in order to make the estimator converge to the correct LDF. %The error should be at most of order $\Delta t$, {\it i.e.}, .  
%As explained in the beginning of Sec.~\ref{subsubsec:CloningRatio}, we perform the cloning procedures at every time-step $dt$, {\it i.e.}~denoting by $\Delta t$ the time interval between cloning steps, we set $\Delta t = d t$.  We then take the $d t \rightarrow 0$ limit to converge to the continuous-time Markov process, which means that the $\Delta t \rightarrow 0$ limit is taken at the same time. 
%
%The $\Delta t \rightarrow 0$ limit was essential in our analysis in order to bound the variation of the population to $\pm 1$ (before rescaling).
%
The error due to non-infinitesimal $\Delta t$ is  at most of order $\Delta t$ as seen from Eq.~(\ref{eq:expected1}) (see also Sec.~\ref{subsubsec:dtDeltat}). 
From a practical point of view,  taking this limit can, however, be problematic, since it requires infinitely many cloning procedures per unit time (as $\Delta t\to 0$). 
%in order to estimate correctly the LDF. 
%
 Interestingly, most of existing algorithms do not take such a limit (see for instance the original version of the algorithm~\cite{giardina_direct_2006}). Empirically, one thus expects that the error goes to zero as $N_c\to\infty$ while keeping $\Delta t$  finite. %, but proving this  remains an open problem. 
%This can be shown that for example, the error can be reduced to smaller order of $\Delta t$, such as $O(\Delta t^d)$ with $d>1$. 
% for the same value of $\Delta t$ but relaxing the condition $\Delta t = d t$.
%
% Within the method developed in this paper, we can formulate the question as follows. Replacing $dt$ by $\Delta t$ in the matrix $\mathcal C$, and we can build a new matrix $\mathcal K^\prime \mathcal C (\mathcal T^{\Delta t/dt})$ and take the $dt \rightarrow 0$ limit while keeping $\Delta t$ non-infinitesimal.  (Note that $\mathcal K^{\prime}$ is not exactly the same as $\mathcal K$ in (\ref{eq:Kexpression}) due to this non-infinitesimal $\Delta t$).  This matrix represents the population dynamics algorithm of a continuous-time process with a finite cloning time-interval $\Delta t$. Empirically, we expect that the error due to this $\Delta t$ goes to zero as $N_c\to\infty$, but proving this remains an open problem. 
%
Within the method developed in this paper, the analytical estimation of this error is challenging (see Sec.~\ref{subsubsec:differenceContinuousTime}) and remains an open problem, but for example, one can approach to this issue numerically at least. 

The second question is about possible extensions of the formulation developed in this paper. 
In our algorithm, we perform a cloning procedure for a fixed time interval, which means that our formulation cannot cover the case of algorithms where $\Delta t$ itself is statistically distributed, as in continuous-time cloning algorithms~\cite{lecomte_numerical_2007}. Moreover, 
 our formulation is limited to Markov systems, although population dynamics algorithms are applied to chaotic deterministic dynamics~\cite{JulienNaturephys_2007,1751-8121-46-25-254002_2013} or to non-Markovian evolutions~\cite{Non-MarkovPD}. Once one removes the  Markov condition in the dynamics, developing analytical approaches becomes more challenging. However, as the physics of those systems are important scientifically and industrially~\cite{PhysRevLett.116.150002}, the understanding of such dynamics cannot be avoided for the further development of population  algorithms.

\begin{acknowledgments}
T.~N.~gratefully acknowledges the  support of Fondation Sciences Math\'ematiques de Paris -- EOTP NEMOT15RPO, PEPS LABS and LAABS Inphyniti CNRS project.
E.~G.~thanks Khashayar Pakdaman for his support and discussions. Special thanks go to the Ecuadorian Government and the Secretar\'ia Nacional de Educaci\'on Superior, Ciencia, Tecnolog\'ia e Innovaci\'on, SENESCYT, for support.
V.~L.~acknowledges support by the National Science Foundation under Grant No.~NSF PHY11-25915 during a stay at KITP, UCSB and
support by the ANR-15-CE40-0020-03 Grant LSD.
V.~L.~acknowledges partial support by the ERC Starting Grant No. 680275 MALIG. 
T.~N.~and V.~L.~are grateful to B.~Derrida and S.~Shiri for discussions. 
We are grateful to S. Shiri for discussions about the first paragraph of Sec.~\ref{subsec:openquestions} (open questions).

\end{acknowledgments}

\bibliography{draft.bib}

\end{document}